\documentclass[conference]{IEEEtran}
\IEEEoverridecommandlockouts
\usepackage{cite}
\usepackage{amsmath,amssymb,amsfonts}
\usepackage{algorithmic}
\usepackage{graphicx}
\usepackage{textcomp}
\usepackage{xcolor}
\usepackage{makecell}
\usepackage{multirow}
\usepackage{enumitem}
\usepackage{float}
\usepackage{algorithmic}
\usepackage{algorithm}
\usepackage{booktabs}

\usepackage{caption}
\usepackage{subcaption}

\def\BibTeX{{\rm B\kern-.05em{\sc i\kern-.025em b}\kern-.08em
    T\kern-.1667em\lower.7ex\hbox{E}\kern-.125emX}}

\begin{document}

\title{\huge Sequential Binary Classification for Intrusion Detection}
\author{\IEEEauthorblockN{
    Shrihari Vasudevan,
    Ishan Chokshi,
    Raaghul Ranganathan
    and
    Nachiappan Sundaram
}
\IEEEauthorblockA{Ericsson, Chennai, India \\
    \{shrihari.vasudevan, r.raaghul and nachiappan.sundaram\}@ericsson.com}
}

\maketitle

\begin{abstract}
Network Intrusion Detection Systems (IDS) have become increasingly important as networks become more vulnerable to new and sophisticated attacks. Machine Learning (ML)-based IDS are increasingly seen as the most effective approach to handle this issue. However, IDS datasets suffer from high class imbalance, which impacts the performance of standard ML models. Different from existing data-driven techniques to handling class imbalance, this paper explores a structural approach to handling class imbalance in multi-class classification (MCC) problems. The proposed approach - Sequential Binary Classification (SBC), is a hierarchical cascade of (regular) binary classifiers. Experiments on benchmark IDS datasets demonstrate that the structural approach to handling class-imbalance, as exemplified by SBC, is a viable approach to handling the issue.\\
\end{abstract}


\section{Introduction}
\label{sec:introduction}

Network IDS monitor network traffic to detect patterns of activity that are unusual or potentially hostile. They can broadly be classified into two types - Signature-based Intrusion Detection Systems (SIDS) and Anomaly-based Intrusion Detection Systems (AIDS) \cite{khraisat2019survey}. While the former is based on identification of matching patterns from a database of previously identified intrusion ``signatures'', the latter is based on establishing a baseline of normal network activity and flagging significant deviations from it, as possible intrusions. SIDS tend to be unable to generalize to previously unseen intrusion ``signatures'' while AIDS tend to have high false-positive rates. The approach proposed in this paper is primarily focused on SIDS, but can also be adapted to provide AIDS capabilities.

Machine Learning (ML)-based IDS are generally more robust and reliable (see \cite{haripriya2018role} and \cite{wang2023robustness}). Standard ML models expect a balanced data distribution between the intrusion (or attack) and non-intrusion events. However, most IDS datasets are comprised of a small number of intrusion events alongside a large volume of normal data; further, different attack types are not equally frequent. Therefore, any ML-based classifier needs to achieve good generalization under severe class imbalance. Other desirable qualities of an effective ML-based IDS solution include the ability to continually update the model with new attack ``signatures'', low latency and interpretability of model decisions. This paper presents an approach to handle class-imbalance in ML-based IDS solutions.

\section{Related Work}
\label{sec:related_work}

Data-driven techniques to handling class imbalance in MCC problems include - (1) resampling techniques, (2) Subagging and (3) the application of sample-weights. Resampling may include under-sampling or over-sampling; the former randomly removes samples from the majority class \cite{devi2020review} whereas the latter creates additional instances of the minority class. Under-sampling may lead to the loss of informative data, negatively impacting training and consequently model performance. Synthetic Minority Oversampling Technique (SMOTE) \cite{chawla2002smote} is a common exemplar of over-sampling techniques. It takes the neighboring data-points and synthetically generates new data using linear interpolation. This approach has been used in ML-based IDS solutions \cite{alfrhan2020smote}, leading to improved outcomes. However, SMOTE can sometimes over-fit on the minority classes or produce data, not representative of the original distribution, resulting in poor generalization (see \cite{lee2021gan} and  \cite{soltanzadeh2021rcsmote}). Subagging \cite{buhlmann2002analyzing} is a computationally cheaper, yet performant, sub-sampling variant of Bagging in ML, where the majority-class data-points are sub-sampled without replacement for the individual classifiers of the ensemble. The application of sample-weights, which may be inversely proportional to their corresponding class-frequencies, is yet another data-driven approach to handling class-imbalance. This paper explores a structural approach to handling class-imbalance. Specifically, it explores if, in a MCC problem, the individual class-frequencies, can be leveraged to handle class-imbalance.

Often, MCC problems on high-dimensional data, have non-linear and complex decision boundaries. The authors of \cite{hastie1997classification} point out that K-class classification rules are easier to learn when one focuses on a single decision boundary at a time. Binarization is a way of converting a K-class problem into a series of 2-class problems, enabling focus on a single decision boundary, at a time. One-vs-All (OVA) \cite{rifkin2004defense} and One-vs-One (OVO) \cite{furnkranz2002round} are popular binarization approaches used in ML. They use a ``divide-and-conquer" approach by constructing multiple binary classifiers, called base-classifiers, and aggregate their results to produce the final outcome. These approaches may enable the learning of smaller and less complex models. However, the computational requirement of OVA increases with larger datasets and classes since all the data-points are required to train every classifier. In the case of OVO, the number of models required to be constructed grows quadratically with the number of classes. The performance of these classifiers is dependent on the aggregation strategy used \cite{zak2020performance}. The algorithm proposed in this paper applies a pairwise and hierarchical binarization technique.

\section{Approach}
\label{sec:approach}

\subsection{Algorithm}
\label{subsec:algo}
\begin{figure}[htb]
\centering
\includegraphics[width=0.8\columnwidth]{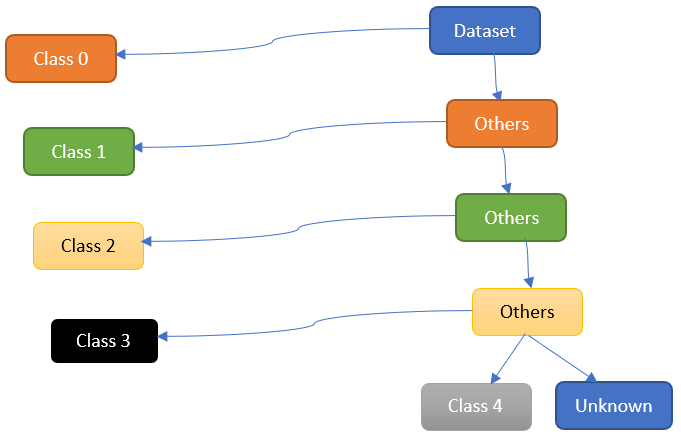}
\caption{Sequential Binary Classification (SBC). Data that are not classified into any of the known classes may be flagged as anomalies or unknown instances.}
\label{fig:fig1}
\vspace{-6mm}
\end{figure}

This paper proposes Sequential Binary Classification (SBC) to handle class-imbalance issues in MCC problems. It is based on a binarization approach that assembles a sequence of binary (base) classifiers, in a hierarchical structure; the approach is referred to as One-vs-All-Others (OVAO) classification. Figure \ref{fig:fig1} depicts an example SBC classifier structure. Classes are first sorted in decreasing order of class-frequency. The first classifier is trained on data which is binarized into the majority-class data-points and those of all remaining classes. Through classification at this stage, the majority-class data-points are removed from contention, and the remaining data-subset is then treated similarly in subsequent classification stages. This process is repeated until all classes are trained. This technique reduces the size of training data for subsequent classifiers. Unlike OVA or OVO, OVAO defines an order in which the individual base classifiers are trained. This enables the model to discriminate between progressively smaller subsets of data. OVAO uses this order to address the issue of increasing model complexity faced by OVA and OVO for large datasets and the increasing number of classes. The algorithm for SBC is summarized below:

\vspace{-1mm}
\begin{algorithm}[] 
     \caption{SBC algorithm}
     \label{algo:sbc}
     \begin{algorithmic}[1]

      \STATE Arrange the classes in decreasing order of frequency and label them from $C_{0}$ to $C_{n-1}$ with $C_{0}$ being assigned to the most frequent class.
      
      \STATE Set $i = 0$
\FOR{$i \gets 0$ \textbf{to} $n-1$}
    \STATE Binarize the class labels, with
       the data-points belonging to class $C_{i}$ as positive and
       the data-points belonging to other classes from $C_{i+1}$
       through $C_{n-1}$ as negative.
    \STATE Select a base classifier to be used for classification and train it on the binarized data subset.
    \STATE Set $i = i + 1$
    \STATE Repeat the above steps until $i = n-1$
    \STATE For the last class, randomly select data from the majority-class ($C_{0}$), or from all other classes ($C_{0}-C_{n-2}$) to be assigned as the negative class for training.
\ENDFOR
     \end{algorithmic}     
\end{algorithm}
\vspace{-2mm}

The ``divide and conquer strategy" used by SBC can help improve classification of minority data-points, by treating multiple minority classes together, thereby reducing class imbalance. The complexity of the boundary is reduced as it is converted into a series of two-class problems, which get progressively smaller. The building of progressively smaller models makes SBC efficient compared to OVA and OVO. The training time and model size would be expected to decrease as we move progressively down the hierarchy since the dataset size keeps shrinking. Model inference follows the same hierarchical structure. Fastest outcomes are obtained for test-instances of class $C_0$ and those of class $C_{n-1}$ would require inference by $n$ classifiers. While this would prove a computational overhead compared to standard MCC for balanced datasets, for test-data reflecting similar class-imbalance as the train-data, this additional overhead may be acceptable to the application context.

\subsection{Hyperparameter Optimization}
\label{sec:sbc_hpo}
Manually defining hyperparameters (HPs) for each base-classifier in SBC will impact its scalability. Additionally, every base-classifier is not guaranteed to have the best performance for the same set of hyperparameters. The base-classifiers need not even be identical. Grid-Search (GS) coupled with cross-validation is a standard approach for hyperparameter optimization (HPO); other effective techniques are based on Bayesian Optimization and Evolutionary methods \cite{bischl2023hyperparameter}. However, GS performs cross-validation on every combination of hyperparameters. This can exponentially increase the time and cost of HPO as the dataset and/or HP search-space size increases. Halving Grid Search (HGS) \cite{pedregosa2011scikit} is a method for speeding up this process. This is an iterative HPO approach where the number of candidate models reduces exponentially with every iteration by comparing its performance with a threshold score of a metric such as accuracy or mean squared error. Simultaneously, the number of data-points available for each candidate model increases at the same rate \cite{li2018hyperband}. HGS-based HPO is used for SBC in this paper.
 
In the context of SBC, HPO using successive halving still consumes significant time, since the number of models required to be trained increases with the number of classes. This paper addresses the issue by bounding the HP search-space using the best HPs from the previous/parent classification stage. The intuition is that models in the SBC hierarchy become progressively simpler due to progressive reduction in data-subset size. Therefore, assuming identical base-classifiers across the SBC hierarchy, a constrained or bounded HP search-space for HPO may suffice. The bound could be a lower or upper bound depending on the HP in consideration. For example, the search-space for the maximum depth of a tree could be expected to become progressively smaller/simpler with increasing level in the SBC hierarchy; it would thus be upper-bounded for training successive SBC stages. This approach is referred to as pruned HGS or pHGS.

\section{Experiments}
\label{sec: exp}

Experiments were conducted to understand the efficacy of (1) the proposed SBC approach, a classifier-structure-based approach to handling class imbalance in MCC problems and (2) the proposed pruned HGS approach to HPO for SBC. Two benchmark IDS datasets, exhibiting severe class imbalance, were used to conduct these experiments.
\begin{itemize}
\item The CICIDS-2017 dataset \cite{sharafaldin2018toward} (CICDS) consists of benign data, 14 attack classes, 78 features, and 2.47 million records. Benign data constituted about 83\% of the dataset. Three dominant attack categories (two denial of service attacks and a port-scan attack) respectively had about 7\%, 5\% and 4\% data. All other attack classes were represented by less than 0.5\% of the data each, with significant imbalance between them as well. The attack-class with the least representation in the dataset had just 11 exemplars. This dataset thus exhibited severe class imbalance. It included duplicate rows, missing, negative and infinity values, and required significant pre-processing leading to a final dataset of 2.2 million records. For experiments in this paper, a 90\%-10\% train-test stratified split was created. 
\item The UNSW-NB15 dataset \cite{moustafa2015unsw} (UNSW) consists of normal and malicious network traffic data created in an emulated environment, for a duration of 31 hours. It consists of nine attack classes, represented by 45 features. Benign data comprised over 36\% of the data, while the top three attack classes were respectively represented by about 23\%, 17\% and 9\% of the data. Other attack classes were represented by about 6\% of the data or less, with significant imbalance between the tail-order classes. The attack class with the least representation had just 174 instances in the entire dataset. A predefined train-test split (68\%-32\%) of the data was provided; these were merged, shuffled and a 90\%-10\% stratified train-test split was created for the experimental objectives of this paper. While no data pre-processing was required for this dataset, it exhibits high inter-class overlap \cite{zoghi2021unswnb15}.
\end{itemize}

Gradient-boosting techniques have arguably become the first-line approach to most MCC problems, due to their superior performance across a range of datasets; XGBoost \cite{chen2016xgboost} is a popular implementation that emphasizes scalability and computational efficiency. To enable a fair comparison and understand the impact of the SBC alone, this paper uses XGBoost, with a Logistic objective function, as the base-classifier (binary) across all stages of the SBC hierarchy and XGBoost with a Softmax objective function, as the baseline MCC classification approach. While Precision, Recall and F1-score are standard MCC performance metrics, for an IDS, the prevention of false-negatives (attack-data deemed benign) is typically as important as the avoidance of false-positives (benign data suspected of being an attack); excessive false-positives undermine the trustworthiness of an IDS and create excessive need for an alternative (often, manual) means of verification. For this reason, the F1-score is used the main measure of performance in this paper. Experiments of this paper report per-class, (simple) average of F1-scores and their standard-deviation. The standard deviation of the F1 scores is indicative of whether improvements in performance are balanced across all classes including minority classes or skewed - for example, improving performance of one set of classes at the expense of others. While several techniques to handle class-imbalance exist, experiments here compare MCC and SBC performance with and without sample-weights - an effective and popular first-line approach to treating class-imbalance, in practice. Similarly, while many methods to hyperparameter optimization (HPO) exist, this paper uses grid-search (GS), Halving Grid Search (HGS) and a pruned variant of HGS proposed in this paper, denoted as pHGS.

\begin{table*}
\centering
\caption{CICDS data - MCC and SBC performance evaluation}
\subcaption*{(MCC - XGBoost, SBC - Sequential Binary Classification, GS - Grid Search, HGS - Halving GS, pHGS - pruned HGS)}
\label{tab:cicds}
\begin{tabular}{lrrrrrrrr}
\toprule
CICDS & MCC + & MCC + & MCC + & SBC + & SBC + & \textbf{SBC +} & SBC + & SBC + \\
Class ID (Name $|$ train size $|$ test size) & GS & GS + & HGS + & GS & GS + & \textbf{HGS +} & HGS & pHGS + \\
Metric & & sample- & sample- & & sample- & \textbf{sample-} & & sample- \\
 & & weights & weights & & weights & \textbf{weights} & & weights \\
\midrule
0 (Benign $|$ 1,844,452 $|$ 204,940) & 1.00 & 1.00 & 1.00 & 1.00 & 1.00 & 1.00 & 1.00 & 1.00 \\
1 (DoS Hulk $|$ 154,798 $|$ 17,200) & 1.00 & 1.00 & 1.00 & 1.00 & 1.00 & 1.00 & 1.00 & 1.00 \\
2 (DDoS $|$ 115,214 $|$ 12,802) & 1.00 & 1.00 & 1.00 & 1.00 & 1.00 & 1.00 & 1.00 & 1.00 \\
3 (Portscan $|$ 81,737 $|$ 9,082) & 0.99 & 0.99 & 0.99 & 0.99 & 0.99 & 0.99 & 0.99 & 0.99 \\
4 (DoS Goldeneye $|$ 9,253 $|$ 1,028) & 1.00 & 0.99 & 0.99 & 0.99 & 0.99 & 0.99 & 1.00 & 0.99 \\
5 (FTP-Patator $|$ 5,340 $|$ 593) & 1.00 & 1.00 & 1.00 & 1.00 & 1.00 & 1.00 & 1.00 & 1.00 \\
6 (DoS slowloris $|$ 4,847 $|$ 538) & 0.99 & 0.98 & 0.98 & 0.99 & 0.99 & 0.99 & 0.99 & 0.99 \\
7 (DoS slowhttptest $|$ 4,704 $|$ 523) & 0.98 & 0.98 & 0.98 & 0.98 & 0.99 & 0.98 & 0.98 & 0.98 \\
8 (SSH-Patator $|$ 2,840 $|$ 315) & 0.98 & 0.97 & 0.97 & 0.98 & 0.98 & 0.98 & 0.98 & 0.98 \\
9 (Bot $|$ 1,754 $|$ 195) & 0.51 & 0.35 & 0.35 & 0.50 & 0.57 & 0.58 & 0.49 & 0.58 \\
10 (Web attack - Brute force $|$ 1,284 $|$ 143) & 0.81 & 0.67 & 0.67 & 0.75 & 0.66 & 0.63 & 0.80 & 0.63 \\
11 (Web attack - XSS $|$ 587 $|$ 65) & 0.26 & 0.35 & 0.35 & 0.42 & 0.53 & 0.52 & 0.40 & 0.53 \\
12 (Infiltration $|$ 32 $|$ 4) & 0.67 & 0.86 & 0.86 & 0.86 & 0.67 & 0.86 & 0.67 & 0.86 \\
13 (Web attack - SQL injection $|$ 18 $|$ 2) & 0.67 & 1.00 & 1.00 & 0.67 & 1.00 & 1.00 & 1.00 & 0.33 \\
14 (Heartbleed $|$ 10 $|$ 1) & 1.00 & 1.00 & 1.00 & 1.00 & 1.00 & 1.00 & 1.00 & 1.00 \\
Accuracy & 1.00 & 1.00 & 1.00 & 1.00 & 1.00 & 1.00 & 1.00 & 1.00 \\
Average F1 & 0.86 & 0.88 & 0.88 & 0.88 & 0.89 & \textbf{0.90} & 0.89 & 0.86 \\
Std-dev F1 & 0.23 & 0.23 & 0.23 & 0.20 & 0.18 & \textbf{0.17} & 0.20 & 0.22 \\
HPO time & 21474.94 & 21702.94 & 4356.63 & 2374.52 & 2445.07 & 881.08 & 959.63 & 541.76 \\
Train time & 132.01 & 140.49 & 121.25 & 22.70 & 26.34 & 21.21 & 19.76 & 16.13 \\
Test time & 1.05 & 0.98 & 1.26 & 61.43 & 65.98 & 60.69 & 61.26 & 59.89 \\
\bottomrule
\end{tabular}
\end{table*}

\begin{table*}
\centering
\caption{UNSW data - MCC and SBC performance evaluation}
\subcaption*{(MCC - XGBoost, SBC - Sequential Binary Classification, GS - Grid Search, HGS - Halving GS, pHGS - pruned HGS)}
\label{tab:unsw}
\begin{tabular}{lrrrrrrrrrr}
\toprule
UNSW & MCC + & MCC + & MCC + & MCC + & \textbf{SBC +} & SBC + & SBC + & \textbf{SBC +} & SBC + & SBC \\
Class ID (Name) & GS & GS + & HGS + & HGS & \textbf{GS} & GS + & HGS + & \textbf{HGS} & pHGS + & pHGS \\
(Train $|$ test size) & & sample-  & sample-  & & & sample-  & sample-  & & sample-  & \\
Metric & & weights  & weights  & & & weights  & weights  & & weights & \\
\midrule
0 (Normal) & 0.93 & 0.89 & 0.89 & 0.93 & 0.94 & 0.93 & 0.92 & 0.93 & 0.92 & 0.93 \\
\hspace{2mm}(83,700 $|$ 9,300) & & & & & & & & & \\
1 (Generic) & 0.99 & 0.99 & 0.99 & 0.99 & 0.99 & 0.99 & 0.99 & 0.99 & 0.99 & 0.99 \\
\hspace{2mm}(52,984 $|$ 5,887) & & & & & & & & & \\
2 (Exploits) & 0.75 & 0.65 & 0.65 & 0.75 & 0.69 & 0.75 & 0.73 & 0.69 & 0.73 & 0.69 \\
\hspace{2mm}(40,072 $|$ 4,453) & & & & & & & & & \\
3 (Fuzzers) & 0.65 & 0.66 & 0.66 & 0.65 & 0.67 & 0.62 & 0.54 & 0.65 & 0.55 & 0.65 \\
\hspace{2mm}(21,821 $|$ 2,425) & & & & & & & & & \\
4 (DoS) & 0.21 & 0.45 & 0.45 & 0.21 & 0.48 & 0.38 & 0.32 & 0.48 & 0.33 & 0.48 \\
\hspace{2mm}(14,718 $|$ 1,635) & & & & & & & & & \\
5 (Reconnaisance) & 0.84 & 0.83 & 0.83 & 0.84 & 0.83 & 0.84 & 0.84 & 0.83 & 0.84 & 0.83 \\
\hspace{2mm}(12,588 $|$ 1,399) & & & & & & & & & \\
6 (Analysis) & 0.19 & 0.14 & 0.14 & 0.19 & 0.23 & 0.20 & 0.14 & 0.22 & 0.15 & 0.22 \\
\hspace{2mm}(2,409 $|$ 268) & & & & & & & & & \\
7 (Backdoor) & 0.18 & 0.08 & 0.08 & 0.18 & 0.19 & 0.19 & 0.16 & 0.20 & 0.15 & 0.20 \\
\hspace{2mm}(2,096 $|$ 233) & & & & & & & & & \\
8 (Shellcode) & 0.65 & 0.55 & 0.55 & 0.65 & 0.61 & 0.63 & 0.57 & 0.57 & 0.58 & 0.59 \\
\hspace{2mm}(1,360 $|$ 151) & & & & & & & & & \\
9 (Worms) & 0.56 & 0.50 & 0.50 & 0.56 & 0.44 & 0.38 & 0.35 & 0.44 & 0.24 & 0.38 \\
\hspace{2mm}(157 $|$ 17) & & & & & & & & & \\
Accuracy & 0.84 & 0.78 & 0.78 & 0.84 & 0.82 & 0.84 & 0.82 & 0.82 & 0.82 & 0.82 \\
Average F1 & 0.60 & 0.58 & 0.58 & 0.60 & \textbf{0.61} & 0.59 & 0.56 & \textbf{0.60} & 0.55 & 0.60 \\
Std-dev F1 & 0.31 & 0.30 & 0.30 & 0.31 & \textbf{0.27} & 0.29 & 0.31 & \textbf{0.27} & 0.32 & 0.28 \\
HPO time & 6488.50 & 6641.60 & 1464.95 & 2456.31 & 2658.89 & 2870.96 & 1079.92 & 1120.17 & 243.06 & 250.88 \\
Train time & 11.56 & 10.38 & 10.03 & 9.79 & 7.94 & 6.33 & 3.69 & 5.01 & 3.38 & 4.42 \\
Test time & 0.08 & 0.08 & 0.07 & 0.08 & 11.80 & 10.08 & 9.28 & 10.39 & 10.11 & 12.20 \\
\bottomrule
\end{tabular}
\end{table*}

\begin{figure*}[htb]
    \centering
    \begin{subfigure}[b]{0.48\textwidth}
        \centering
        \caption{MCC with grid-search (GS)-based HPO}
        \label{fig:mcc_gs}
        \includegraphics[width=\textwidth]{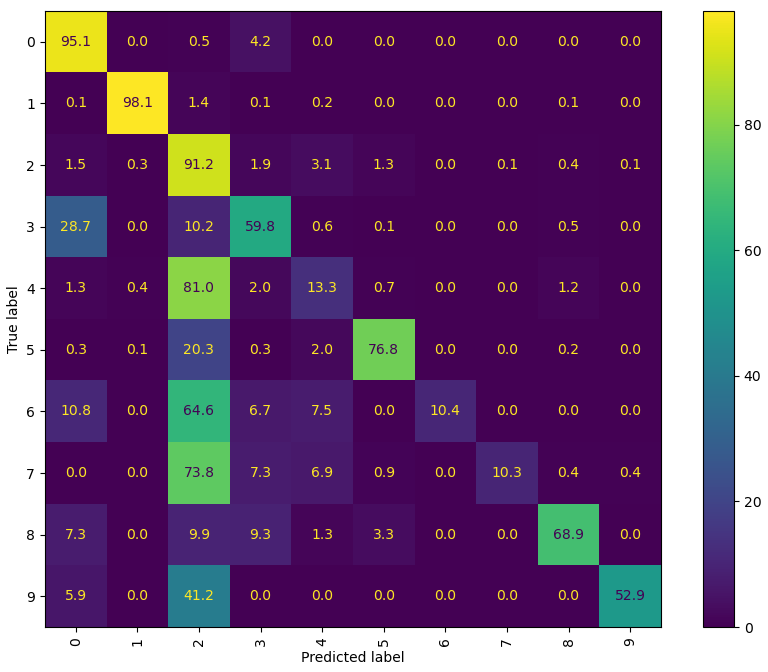}
    \end{subfigure}
    \begin{subfigure}[b]{0.48\textwidth}
        \centering
        \caption{MCC with grid-search (GS)-based HPO and sample-weights}
        \label{fig:mcc_gs_sw}
        \includegraphics[width=\textwidth]{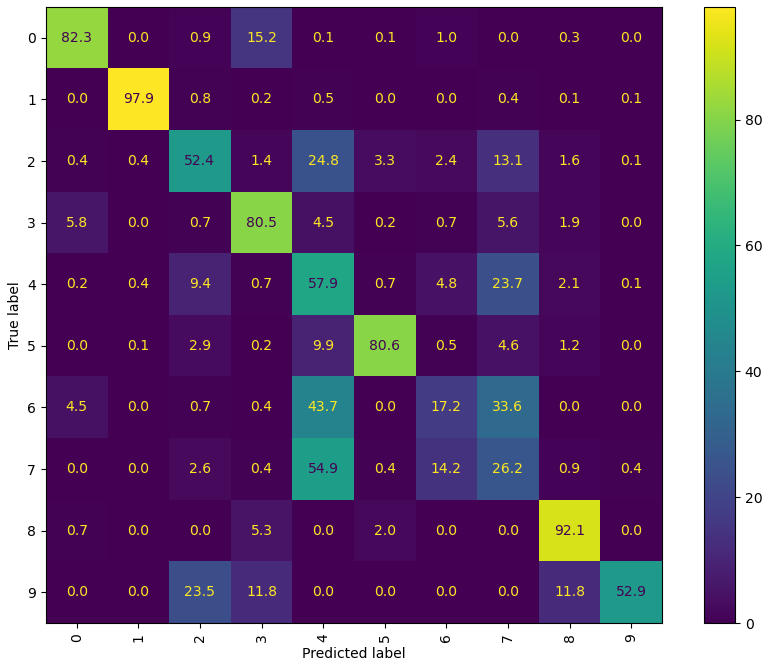}
    \end{subfigure}\\
    \begin{subfigure}[b]{0.48\textwidth}
        \centering
        \caption{SBC with grid-search (GS)-based HPO}
        \label{fig:sbc_gs}
        \includegraphics[width=\textwidth]{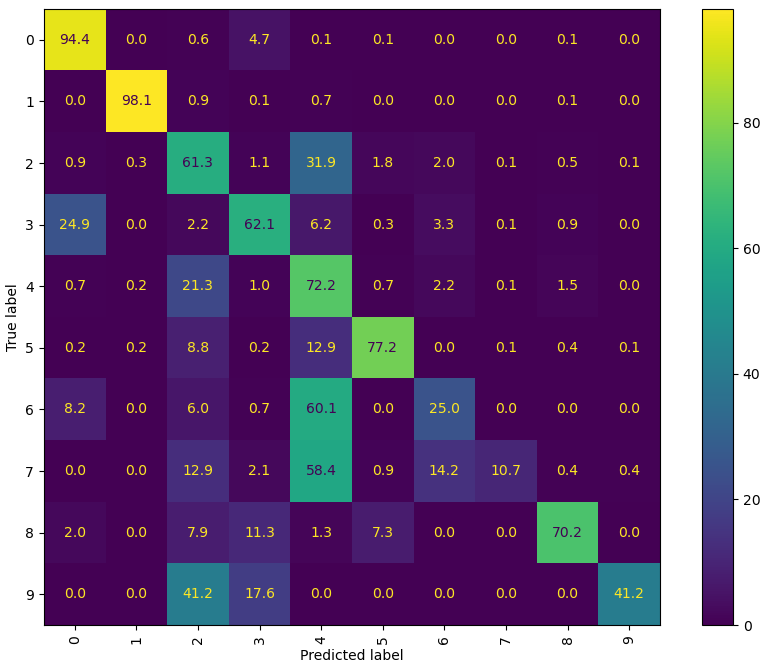}    
    \end{subfigure}
    \begin{subfigure}[b]{0.48\textwidth}
        \centering
        \caption{SBC with grid-search (GS)-based HPO and sample-weights}
        \label{fig:sbc_gs_sw}
        \includegraphics[width=\textwidth]{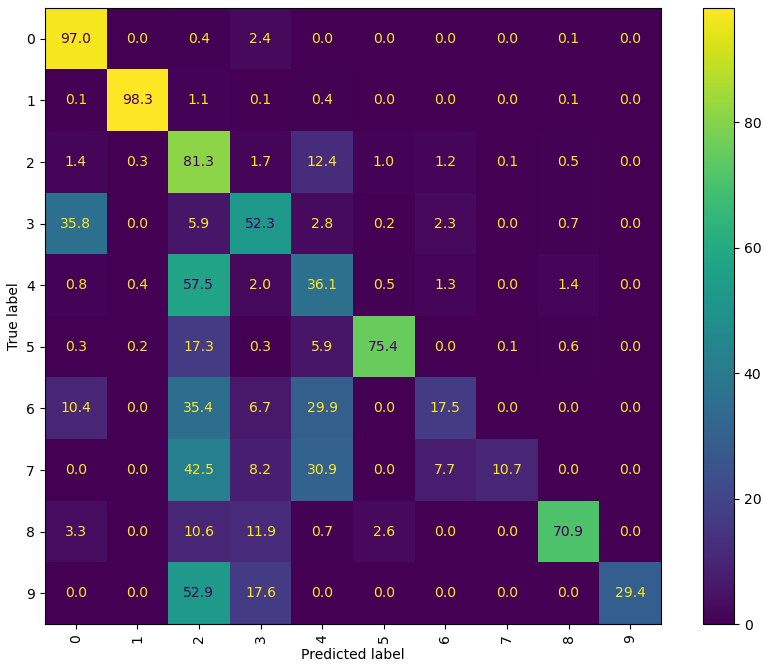}
    \end{subfigure}
    \caption{UNSW data-set - Normalized (\%) confusion matrices for MCC vs SBC with and without sample-weights}
    \label{fig:unsw}
    \vspace{-5mm}
\end{figure*}

Tables \ref{tab:cicds} and \ref{tab:unsw} show the performance of different MCC and SBC approaches, tested on the CICDS and UNSW datasets respectively. The CICDS dataset, while large, showed relatively discernible patterns, listed below.
\begin{itemize}
\item Classes 9-14 were the minority classes, the tail-end of a severe class-imbalance problem with class 9 having 1000x less data than class 0 (benign instances). Differences between methods were evidenced in these classes; in classes 0-8 and 14, performance were near identical across methods. MCC with GS-based HPO and no class-imbalance treatment via sample-weights, provided the baseline.
\item The best outcome was obtained using SBC, optimized using HGS and with sample-weights. This approach produced higher average F1 and lower standard-deviation of F1 scores across all classes, including the minority classes. The former indicates better overall performance and the latter is suggestive of balanced improvement across all classes. The approach uses both structure and sample-weight approaches to treating class imbalance. 
\item MCC approaches produce lower average F1 and higher standard-deviation of F1 scores, compared to all SBC approaches except the one that used pHGS for HPO.
\item Comparing SBC using GS-based HPO with MCC using GS-based HPO and sample-weights to treat class-imbalance, in all but one minority class (13), SBC, using structure alone, improves (or matches) performance over MCC, using sample-weights. A similar trend is observed for HGS-based HPO.
\item Grid-search (GS) is far more expensive than HGS or pHGS. This is expected and intuitive. Performance obtained with HGS-based HPO is competitive with those of GS. Pruning in pHGS results in reduction in HPO time, but accompanied with a drop in performance.
\item SBC dramatically reduces HPO and training times compared to MCC. On the same note however, SBC has significantly higher inference times, compared to MCC. Both trends are attributable to the binary and sequential nature of SBC. In this dataset, SBC inference was able to process about 4000 data per second, whereas MCC processed about 247000 data per second.
\end{itemize}

Next, in the case of the UNSW dataset, the following patterns were observed from Table \ref{tab:unsw} and Figure \ref{fig:unsw}. 
\begin{itemize}
\item This dataset has both significant class imbalance and significant inter-class overlap as evidenced by Figure \ref{fig:mcc_gs}. The classes which differentiate performance among the tested methods include 0, 2, 4, 6, 7, 8 and 9; of these, the last four are minority classes. Most mis-classifications occur with class 2 i.e., most instances of other classes are mis-classified as being of class 2. To a lesser extent, this issue occurs with classes 0 (normal exemplars), 3 and 4.

\item Looking at the performance of the baseline (MCC with GS-based HPO) and of that with sample-weights to treat class imbalance, Figures \ref{fig:mcc_gs} and \ref{fig:mcc_gs_sw} suggest that outcomes of minority classes improve with more correct classifications, however, those of majority classes drop. The pattern of most data being incorrectly mis-classified as class 2 or 0, now, also incorrectly, shifts to classes 4 and 7 and to a smaller extent, classes 3 and 6. This suggests a tendency to reduce false-negatives at the cost of increasing false-positives. The net result (aggregate and standard deviation of F1 scores) are therefore similar. A similar trend is observed with HGS-based HPO, seen in Table \ref{tab:unsw}; GS based HPO however takes significantly longer than HGS.

\item Considering the baseline MCC outcomes in Figure \ref{fig:mcc_gs} and those of SBC with GS-based HPO (Figure \ref{fig:sbc_gs}), it was observed that even in SBC, the pattern of mis-classification shifts, but to a smaller extent than MCC. Outcomes mis-classified as class 2 in MCC get mis-classified as class 4 in SBC, and to a smaller extent, as classes 0, 2, 3 and 6. Outcomes misclassified as class 0 (normal) come down significantly, but to a lesser extent than with MCC coupled with sample-weights in Figure \ref{fig:mcc_gs_sw}. Majority class performance in SBC (Figure \ref{fig:sbc_gs}) does not drop as significantly as in the case of MCC with sample weights (Figure \ref{fig:mcc_gs_sw}).

\item The previous observations suggest that when data has significant inter-class overlap, adding sample-weights to MCC tends to improve minority class performance, possibly, at the expense of majority class performance. A strong drop in false-negatives tends to be accompanied with a strong rise in false-positives. In comparison, SBC, using a structural approach to handle class imbalance, is able to balance outcomes across classes - majority class performance is better maintained, while minority class performance also improves. This is also indicated by a smaller standard deviation of F1-scores (see Table \ref{tab:unsw}) while a similar average of F1-scores indicates that SBC is competitive. SBC tends to better balance the reduction of false-negatives with an increase in false-positives.

\item Looking at all columns in Table \ref{tab:unsw}, the best SBC outcome (with GS-based HPO and no sample-weights) was marginally better than the best MCC outcome (for this dataset, the baseline) in terms of aggregate F1-score but significantly better than every MCC outcome in terms of standard-deviation of F1-scores across individual classes.

\item Based on the columns, in Table \ref{tab:unsw}, showing SBC performance with and without sample-weights, using either HGS or GS-based HPO, it was inferred that given the inter-class overlap in the UNSW dataset, the addition of sample-weights to SBC results in a net performance drop, because the weights exacerbate the mis-classifications that may be attributable to inter-class overlap. Figures \ref{fig:sbc_gs} (SBC), \ref{fig:sbc_gs_sw} (SBC with sample weights) and \ref{fig:mcc_gs} (MCC baseline) show that adding sample weights to SBC, degrades performance and reverse-shifts the mis-classifications back towards the baseline outcome of MCC without any treatment for class imbalance. 

\item Like in the CICDS dataset, SBC HPO and train times were significantly lower than those of MCC but SBC inference times were much higher than those of MCC. Pruning in HGS reduced HPO time significantly; however, in this case and unlike in CICDS, outcomes remain competitive with that of HGS without pruning. 
\end{itemize}

Future work will explore additional datasets, threshold-selection and compare with other class-imbalance methods. 

\section{Conclusion}
This paper presented Sequential Binary Classification (SBC) as a structural approach, of one-vs-all-others (OVAO) binarization, to address the issue of class-imbalance in multi-class classification problems. Experiments suggest that a structure-based approach to handling class imbalance, as evidenced by SBC, is a viable and effective alternative to an established approach like the use of sample-weights. For datasets with significant inter-class overlap, SBC is best used as an exclusive approach, and not coupled with sample-weights. While HPO and training times are reduced, inference times increase significantly; this needs to be considered in a potential application context. Further, SBC provides flexibility in the use of different (e.g., simpler) and even non-tree base-classifiers across different stages.

\bibliography{references}
\bibliographystyle{ieeetr}
\end{document}